# The Separation and H-alpha Contrasts of Massive Accreting Planets in the Gaps of Transitional Disks: Predicted H-alpha Protoplanet Yields for Adaptive Optics Surveys

Laird M. Close[1][2]

## ABSTRACT

We present a massive accreting gap (MAG) planet model that ensures large gaps in transitional disks are kept dust free by the scattering action of three co-planar quasi-circular planets in a 1:2:4 Mean Motion Resonance (MMR). This model uses the constraint of the observed gap size, and the dust-free nature of the gap, to determine within ~10% the possible orbits for 3 massive planets in an MMR. Calculated orbits are consistent with the observed orbits and H$\alpha$ emission (the brightest line to observe these planets) for LkCa 15 b and PDS 70 b and PDS 70 c within observational errors. Moreover, the model suggests that the scarcity of detected H$\alpha$ planets is likely a selection effect of the current limitations of non-coronagraphic, low (<10%) Strehl, H$\alpha$ imaging with Adaptive Optics (AO) systems used in past H$\alpha$ surveys. We predict that as higher Strehl AO systems (with high-performance custom coronagraphs; like 6.5-m Magellan Telescope MagAO-X system) are utilized at H$\alpha$ the number of detected gap planets will substantially increase by more than tenfold. For example, we show that >25±5 new H$\alpha$ "gap planets" are potentially discoverable by a survey of the best 19 transitional disks with MagAO-X. Detections of these accreting protoplanets will significantly improve our understanding of planet formation, planet growth and accretion, solar system architectures, and planet disk interactions.

[1] Corresponding author lclose@as.arizona.edu
[2] Center for Astronomical Adaptive Optics, Department of Astronomy, University of Arizona, 933 N. Cherry Ave. Tucson, AZ 85718, USA

# 1. INTRODUCTION

It is now well observationally established that some gas giant protoplanets, such as PDS 70 b and PDS 70 c, pass through a period of high luminosity as they accrete hydrogen gas from their circumplanetary disks producing detectable H$\alpha$ emission. This was most clearly demonstrated in the direct detections of H$\alpha$ emission from LkCa 15 b (Sallum et al. 2015), PDS 70 b (Wagner et al. 2018), and PDS 70 c (Haffert et al. 2019). Direct observations of protoplanets (defined here as accreting planets) are a key window into this poorly understood process of planet formation and accretion from a circumplanetary disk. While the exact mechanisms of planetary accretion are not yet fully understood, massive planets could magnetospherically accrete, via magnetic fields, directly onto the polar regions of the planet (Zhu et al. 2016; Thanathibodee et al. 2019 and references within).

## 1.1. A Lack of H$\alpha$ Protoplanets?

Protoplanets can be difficult to discover since young, gas-rich, (<10 Myr) stars are D>120pc from Earth. Therefore, we must look to young stars with protoplanets on relatively rare wide ($\geq$ 10 au; $\geq$80mas) orbits. But how do we find such stars? Selection of optimal targets is facilitated by ALMA dust continuum surveys like DSHARP (Andrews et al. 2018), which have identified many young stars with transitional disks and wide disk gaps. For example, an exhaustive archival search of the ALMA archive by Francis & Van Der Marel (2020; hereafter FVDM) find 38 young stars with well resolved wide 20-80 au gaps (see Fig 1). These dust free gaps, or cavities, can be best explained by having multiple protoplanets clearing the gap. Dodson-Robinson & Salyk (2011) showed hydrodynamical simulations of single planet at 10 au only being able to clear a ~7 au wide gap (roughly ±4 Hill Spheres), but three 3$M_{jup}$ "gap planets" inside a 20 au gap could maintain a dust free gap.

In section 2 we motivate this study and add background information. In section 3 we present a new "gap planet" model and in section 4 project it onto 9 of the most popular large gap disks already observed with AO for H$\alpha$ protoplanets. In section 4 we also project our model on the best 23 large gap transitional disks of FVDM. We use this model to predict what the detctable population of gap planets might be. In section 5 we discuss these results and what the success of more sensitive future Hα searches could be. In section 6 we state our conclusions.

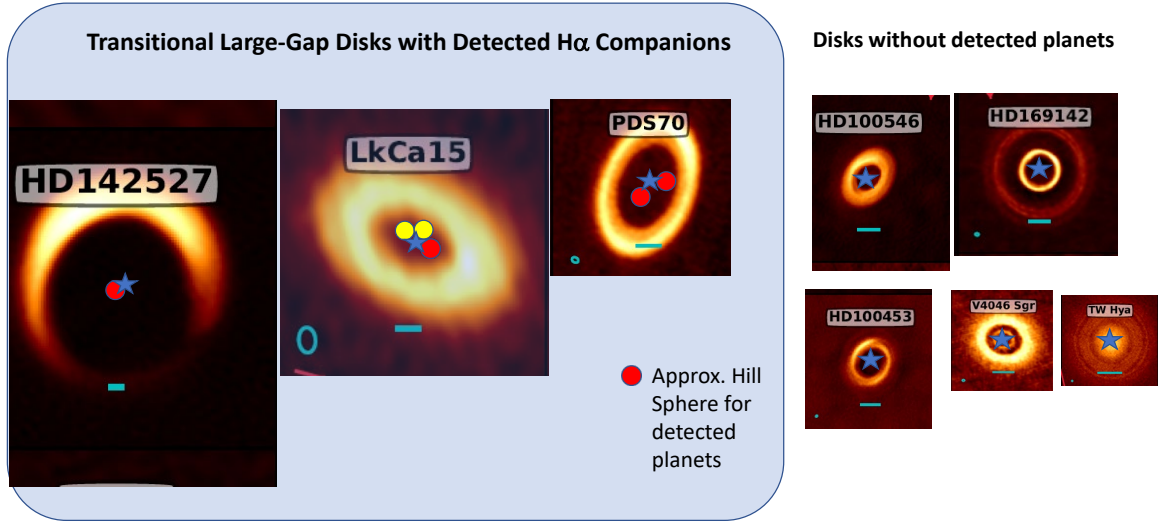

**Fig. 1:** Big ALMA cavities yield gap planets. Early Hα SDI results include the exciting detections of HD142527B (Close et al. 2014), LkCa 15b (Sallum et al. 2015) and PDS 70b (Wagner et al. 2018) and PDS 70c (Haffert et al. 2019). These images are from ALMA reproduced, with permission, from FVDM. We note that each image has been scaled so that the 30 AU blue horizontal bar is the same size, highlighting correct relative sizes (except for HD142527 which is too large).

## 2.0 MOTIVATION & BACKGROUND

### 2.1. Reasons Protoplanets Might be Faint at Hα --or is it all a Selection Effect?

As a result of the SPHERE/ZIMPOL non-detections of new Hα planets Cugno et al. (2019) and Zurlo et al. (2020) there has been a series of papers describing why Hα planets might be apparently so rare. Brittain et al. (2020) suggest that planetary accretion could be episodic in nature similar to an "FU Ori" type of outburst. Hence, it could be hard to catch the planets when they are near their peak of accretion/Hα luminosity. Brittain et al (2020) suggest that PDS 70 b and c are in the middle between the quiescent and burst state. If either planet dramatically increases its Hα brightness in the future this would prove the theory of Brittain et al. (2020). In another recent study, detailed three dimensional thermo-hydrodynamical simulations of Szulagyi & Ercolano (2020) show that the extinction from dust could extinguish Hα from all but the most massive ($\geq 10 M_{jup}$) planets. However, given that the masses of the PDS 70 c and b planets are ~2 and ~4 $M_{jup}$ (respectively; Wang et al. 2020), then the dust free "gas-only" models of Szulagyi & Ercolano (2020) are the only models in that study that can explain the observed properties and line strengths of PDS 70 b and c. In fact, a detailed physical model of magnetospherical accretion by Thanathibodee et al. (2019) shows that the accretion of PDS 70 b is well explained by magnetospherical accretion but the efficiency of H$\alpha$ line luminosity productivity falls dramatically if the mass accretion rate falls below a certain crossing point.

There is currently a tension in the literature as to how bright such "gap planets" should be at H$\alpha$. A key question that has not been rigorously poised or answered is: just how many of these H$\alpha$ gap planets should we have detected already with current AO sensitivities? Are the null results (save PDS 70 b and PDS 70 c and LkCa 15 b) significant --or simply a selection effect of the limits of the AO surveys themselves? These AO systems cannot correct the atmosphere very well at H$\alpha$ (656.3nm is quite blue for AO correction; Close et al. 2018), particularly with fainter guide stars, since the coherence patch size ($r_o$) of the atmosphere $r_o=15(\lambda/0.55)^{6/5}$ cm on a 0.75" seeing night. Hence it is clear that $r_o$ at H band is 56cm but at H$\alpha$ is just 18cm. So, only AO systems with <18cm sampling of the telescope primary mirror will make the highest contrast images at H$\alpha$ (see Close 2016 and Close et al. 2018 for reviews). For example, the Strehl of the corrected wavefront may be 75% (residual wavefront error 140nm rms) at H band (1656nm; where SPHERE was designed to work) but at H$\alpha$ it is just 16% --so 84% of the starlight is outside of the diffraction PSF and is swamping any H$\alpha$ light from the planets. This simple scaling has another "hit" for H$\alpha$ contrasts, the Strehls are so low that no coronagraph is used in any of the datasets of Cugno et al. (2019) and Zurlo et al. (2020), hence the inner 0.2" of the SPHERE/ZIMPOL images have 100% of the diffracted and atmospheric speckles swapping the individual images (making contrasts of $\leq 10^{-4}$ at 100mas impossible). Similar limits apply to MagAO's H$\alpha$ imaging as well (no coronagraph, low Strehls). All these effects are not trivial and have made it difficult for MagAO to detect PDS 70 b even at 4$\sigma$ and impossible for SPHERE/ZIMPOL for PDS 70b (or c) at H$\alpha$ (Wagner et al. 2018). The 4$\sigma$ detection of PDS 70 b by MagAO was confirmed by VLT/MUSE by Haffert et al. (2019). But in all these cases, the detections were difficult and required good correction on PDS 70 which is not a very bright guide star at R~11.8 mag. Hence, it is fair to ask: Is it a selection effect that, to date, most of the well AO observed transitional disks are around the brightest targets which, in turn, may very well be the "worst" targets since the high luminosity of the central stars drown out the H$\alpha$ planets? We first need a general predictive model of gap planets contrasts and separations to start to answer these valid questions about AO observational selection effects and planet yields.

In this paper we model a complete population of gap planet masses and orbits based purely on the physical characteristics of the ALMA observations of these transitional disks (size of gap, inclination of the gap, the stellar mass, and stellar R mag). The model creates a set of 3 planets in each of these large-gap disks that explains the dust-free gaps yet also stay dynamically stable >10Myr. To model the H$\alpha$ line luminosity of these planets, we then use a power-law break in the efficiency of H$\alpha$ luminosity based on planetary mass accretion rates motivated by the physical planetary accretion models of Thanathibodee et al. (2019). We then derive relations between the brightness of the planet/star at H$\alpha$ (H$\alpha$ contrast; hereafter: $\Delta$magH$\alpha$). Allowing the prediction of the H$\alpha$ contrast and the angular separation, of a synthetic population of gap planets. Once this population is compared to average sensitivity curves of MagAO, SPHERE, and MUSE (all based on the detection sensitivity of real H$\alpha$ planets) we can test whether the recent lack of detections prove H$\alpha$ planets are truly very faint. We may also find that there is a large population of gap planets that have not yet been discovered simply due to selection effects of the current surveys.

### 2.2. Current Approaches to Imaging H-alpha Planets

We can now make deep images with spatial resolutions of 20-25 mas (0.025") (Close et al. 2013; Close et al. 2014b; see review Close 2016; Close et al. 2018) from the ground at H$\alpha$. Previously with the first large $\geq$6.5m visible AO system (MagAO) we achieved contrasts of $10^{-3}$ at 0.2" arcsec from a bright star (Wagner et al. 2018) as we detected the in-fall of hydrogen gas as it accreted onto low-mass companions in the cleared gaps of transitional disks (Close et al. 2014; Sallum et al. 2015; Wagner et al. 2018; Haffert et al. 2019). All these detections were made possible by the Simultaneous Differential Imaging (SDI) technique where an H$\alpha$ image has a scaled continuum image subtracted from it to remove some of the PSF speckles and reveal any H$\alpha$ planets (Close et al. 2004; 2014).

### 2.3. New H$\alpha$ detection Techniques: Extreme Visible AO (MagAO-X) and MUSE

These impressive "H$\alpha$ AO" detections were done with older AO systems (VLT/SPHERE, VLT/MUSE, Magellan/MagAO) with relatively low (<10%) Strehls at H$\alpha$. However, we have just had first light with the world's newest extreme AO system MagAO-X. MagAO-X is unique --it was designed from the start to work in the visible at high Strehl (Males et al. 2018; Close et al. 2018). MagAO-X yields a superior level of wavefront control with a 2040 actuator Tweeter deformable mirror (DM) and a unique "extra" DM to eliminate all Non-Common Path (NCP) errors between the science and wavefront sensing channels, minimizing coronagraphic leak. Wavefront sensing (WFS) with MagAO-X's very low noise (<0.6 rms e- read noise) EMCCD pyramid WFS detector allows Strehls of 50% to be obtained while closed loop at 2kHz (residual WFE <120nm rms – as demonstrated on-sky (Males et al. 2020, in prep). The low noise of this sensor allows good correction even on faint R~13 mag guide stars in good 0.5" seeing conditions. The MagAO-X system with up to 1500 corrected modes maps to ~15 cm/actuator, making it the highest sampled AO system in the world. So deeper, much more sensitive surveys for H$\alpha$ planets are finally possible.

Two different approaches could lead to substantial increases in the number of H$\alpha$ planets detected. For bright (R<12mag) targets, planets could be detected with MagAO-X and for those fainter (R>12mag) with the laser guide star fed VLT/MUSE IFU. Hence, it is very important for the future of this field to know if the current lack of H$\alpha$ detections is fundamental to the H$\alpha$ line luminosity production (and extinction) mechanisms --or simply a result of selection effects in the current generation of AO surveys.

## 2.4. Past Successes Detecting Gap Protoplanets

### 2.4.1. The First H$\alpha$ Planet: LkCa 15 b

It has been known for some time that there is a special class of circumstellar disk called a transitional disk. These disks have a lack of NIR and MID-IR flux in their SEDs (Espaillet et al. 2008) which when imaged by sub-mm interferometry revealed large "gaps" (hole or cavity) in the dust distribution (see Fig 1). An excellent example of such a disk is that of LkCa 15 (Andrews et al. 2016). LkCa 15 had risen to importance since the speckle masking work on the 10m Keck II telescope of Kraus & Ireland (2012), found evidence of a companion inside the large (~72 au) gap transitional disk around LkCa 15. Later NRM masking work on the LBT telescope and re-reduction of the Keck data suggested that there were 3 objects in the LkCa 15 gap (Sallum et al. 2015), moving with a clear orbital motion over a 5 year period. Simultaneously, yet independently of the LBT work, MagAO H$\alpha$ observations found a 6$\sigma$ H$\alpha$ point source coincident with the NRM position for the LkCa 15 b planet on a 15 au orbit (0.1") around LkCa 15 A. Only LkCa 15 b was detected in H$\alpha$. LkCa 15 (at DEC=+22º and V~12) proved to be a challenging target from the south and no following MagAO datasets were of the 0.5" seeing quality of the published one. Recently Currie et al. (2019) have claimed that the NRM results could be due to bright dust clumps in the inner disk. However, FVDM find no evidence of a significant inner dust in orbit around LkCa 15 in deep ALMA continuum images, moreover recent high-resolution images from ALMA by Facchini et al. (2020) estimate that the compact optically thick central disk around LkCa 15 cannot be greater than ~0.3 au. Hence these "clumps" (unseen by ALMA) could be compact and optically thick at sub-mm wavelengths and could, therefore, be compact circumplanetary disks around planets b and c, which explains why the dust was difficult to detect by ALMA (Wu et al. 2017b). In any case, the observed strong Hα emission from LkCa 15b cannot be explained by anything other than a protoplanet. However, LkCa 15 b still needs a significant detection at another epoch at Hα to be 100% unambiguous.

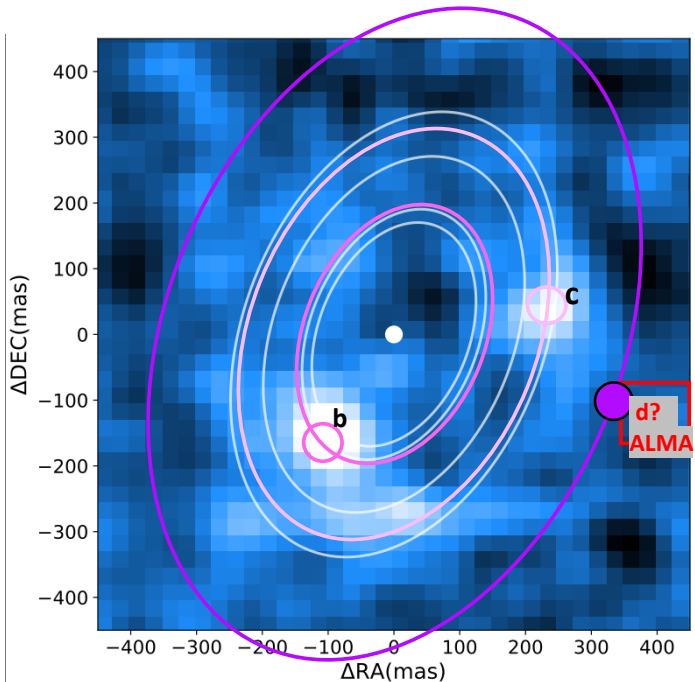

**Fig 2:** Hα image of PDS70b, c (Haffert et al. 2019). The predicted inner ($a_1$=0.30Rcav) and middle ($a_2$=0.47Rcav) orbits from our MAG model applied to the Rcav=74au PDS70 cavity/gap. The open circles show the predicted planet positions (PA is fit to the planets). Excellent agreement with observation. Planet "d" is hinted at by an ALMA detection (red box). We can't detect outer planet d at Hα due to the disk's inclination (see red line in Fig 3).

### 2.4.2. PDS 70 b and c

As figure 1 shows there are a handful of very large gaps known. It is interesting to note that in *every* case of a very wide gap there was an accreting companion discovered inside the gap. An excellent example is the PDS 70 system. PDS 70 A is a 0.8$M_{sun}$ TTauri star of age 5Myr accreting at ~6x10$^{-11}M_{sun}$/yr (Thanathibodee et al. 2020), which has a spectacularly large 74 au wide gap (see Fig 1). Imaging with SPHERE was able to discover thermal emission from the circumplanetary disk and atmosphere from the gap planet PDS 70b (Keppler et al. 2018). We were able to use MagAO to discover Hα likely from magnetospherical accretion onto PDS 70b (Wagner et al. 2018). The VLT's MUSE IFU was used to confirm the H$\alpha$ emission from PDS 70b and discovered PDS 70c as another H$\alpha$ protoplanet inside the gap (Haffert et al. 2019; see Fig 2). Since, the separations of PDS 70b and c are rather large (~0.19" and ~0.23" respectively), telescopes like Keck at L' (3.8 µm) are able to follow-up these planets to detect their circumplanetary disk emission (Eisner 2015) where the masses are measured to be roughly ~2-4$M_{jup}$ for b and ~1-2$M_{jup}$ for c (Wang et al. 2020).

We caution that while L' is superior to H$\alpha$ to piercing any dust extinction, it can be hard to achieve the required spatial resolutions at longer wavelengths. For example, PDS 70 b at 0.185" translates to just 2.5 $\lambda/D$ at L' with the large D=10m Keck telescope, this is very close to the inner working angle limit for high-contrast (~$10^{-4}$) direct detection. Closer-in planets at, say, ~0.1" (1.3$\lambda/D$ at Keck) would really require an ELT sized aperture for high-contrast direct direction at L'. In contrast, H$\alpha$ is a 5.8x shorter wavelength, so even a smaller D=6.5m telescope finds a 0.1" planet at 5$\lambda/D$ at H$\alpha$, and so can be detected, quite easily, (especially if there is use of a coronagraph) even if it is at $10^{-4}$ contrasts.

*2.5. How Transitional Disks Can Maintain These Large Dust Free Gaps*

In transitional disks, whose dust cleared central cavities are optically thin, there is little extinction towards the protoplanet. This is especially true for the polar regions of the planet where the H$\alpha$ is created by the shock from the magnetospherical accretion. In fact, it has been theorized that the reason these gaps stay dust free is due to the sculpting influence of giant planets (Alexander & Armitage 2009). The best gap planet candidate systems are the so-called "wide-gap" transitional disks that may need multiple >1 $M_{jup}$ mass planets to keep the gap cleared, since these gaps are ±4$R_H$ (Hill spheres) where the size of any one planet's Hill sphere:

$$R_H = (33/40)(a/10)(M_p/M_*)^{1/3} \text{ au;} \qquad (1)$$

where $M_p$ is the mass of the planet in $M_{jup}$ and $M_*$ is the mass of the star in solar masses. Hence even a massive 5 $M_{jup}$ planet can only open a 10±5 au dust free gap at *a*=10 au in 10 Myr (according to the hydrodynamical (HD) simulations of Dodson-Robinson & Salyk (2011); in agreement with the more recent 3-D HD models of Sanchis et al. 2020). The popular "gap planets" theory of Dodson-Robinson & Salyk (2011) makes a convincing case that multiple (3) massive (3 $M_{jup}$), co-planar, gas planets can produce all of the commonly observed properties of transitional disks, including large gaps. Such gap planets scatter the dust creating the observed gaps --but let the gas pass through the gap. Some gas accretes onto the planets and the rest onto the star. This allows for long lived (>5 Myr) gaps around continuously accreting TTauri stars. These timescales fit the observations (Fig. 1,2) much better than the competitor "photoevaporation" theory where once a large gap is cleared, the theory predicts accretion onto the star quickly ends. There are too many large gaps over a large age range (1-10 Myr) for photoevaporation to explain all these features (see the review of Owen 2016 and references within). But we should also be open to the question: if gap planets clear all these cavities, why have we not observed H$\alpha$ planets in *all* of these cavities (Brittain et al. 2020)? In the following sections we will try to address that question.

One thing appears clear, that in dust-free transitional disks there should not be a large extinction along the observed line of sight to the planet (these are, after all, largely dust free gaps). By targeting transitional disks that are not exactly edge-on (see Fig. 1 for examples). We should be able to directly detect planetary H$\alpha$ emission produced by gas accreting onto the planet (onto some shock surface/boundary) without high extinction along the line of sight. This exact point is proven by the 3-D HD simulations of Sanchis et al. (2020) who find that once a significant gap is opened by a massive planet the amount of residual dust leads to <1 mag of extinction towards the planet at the wavelength of H$\alpha$.

## 3. A NEW MODEL FOR MASSIVE ACCRETING GAP PLANETS

### *3.1 The MAG Model*

The model of Dodson-Robinson & Salyk (2011) (herein the DRS model) has continued to be the most common explanation for these large gaps in the literature. However, there are a few inconsistencies with observational data. For example, DRS predicts that all the gap planets have the same mass, implying that all gap planets would be equally bright at H$\alpha$ and other wavelengths. Imaged multi-planet systems are very rare (only 3 systems known) yet there is a uniform trend that the outer planet is always lower mass than inner planets. For example, PDS 70 b (inner planet) is roughly 2x the mass of outer planet PDS 70 c (Wang et al. 2020), similarly HR8799 c (middle planet) is roughly 1.4x the mass of HR8799 b (outer planet). Moreover, the mass of the inner planet HR87799 e is 1.4x the mass of HR8799 d (Marios et al. 2010; Currie et al. 2011). Moreover, the newest addition to this small list of wide multi-planet systems TYC 8998-760-1 also has an inner planet 1.75x the mass of the outer planet (Bohn et al. 2020). So, there is a well observed steady trend of massive, wide, multi-planet systems increasing in mass as you approach the inner most planet. Even in our own Solar System Saturn is 3.0x more massive than the outer planets Uranus plus Neptune, and Jupiter is 3.3x more massive than Saturn. Simply taking the average of these implies that middle planet should be 1.4-3.0 more massive than the outer planet, and the inner planet also 1.4-3.0 more massive than the middle planet. To be consistent with this observed trend we can average the above mass ratios (2.0, 1.4, 1.75, 3.0)=<2.0> and conclude that, on average, the middle planet is 2x the mass of the outer planet and the inner planet is 2x more massive than the middle planet.

Also, the DRS model claims there must be a large massive planet within $4R_H$ of the cavity edge (Rcav). In fact, detailed modeling by Dong & Fung (2017) showed that disk edge profiles (like those in Fig 1) showed signs of being sculpted by outer planets of masses no more than ~1 $M_{jup}$. Adding additional observational evidence that the outer planet is lower mass then the 3 $M_{jup}$ suggested by DRS.

Also, the long term stability of the DRS model was questionable past 10 Myr given that the planets were all massive but were not in a stable mean motion resonance (MMR). Nature prefers an MMR configuration for stability with massive planets. Our best example of a massive wide multiplanet system is the HR8799e,d,c,b set of 4 massive (5-10 $M_{jup}$) planets spanning over 70 au. We have been observing this system long enough to be able to fit HR8799 with a classic 1:2:4:8 MMR that allows long term stability (see Gozdziewski & Migaszewski 2014).

The PDS 70 b and c planets have been shown to be in a stable >1Myr 1:2 MMR already in the 2D HD models of Bae et al. (2019). This gives us a strong indication that MMRs do play an important role in maintaining the long term stability of having multiple massive planets inside large gaps. We can use the strong orbital relationship between planets in an MMR to synthesize a large population of gap planets each custom fit to the cavity it is resident in. Recent N-body simulations of Wang et al. (2020b) show that in a gapped accretion disk it is possible to have 3 massive gap planets be stable well past the disk dissipation time out to >10 Gyr if the planets are all very close to an MMR --near-resonance but not perfectly locked into one. A case of near-resonance orbits will have semi-major axes close enough to a perfect MMR that it will not significantly affect our MAG model predicted separations on sky, so we adopt a perfect MMR for simplicity -- while being cognizant that near-resonance systems would also look very similar and may be preferred in some cases by nature.

We can build on strengths of the DRS model with a new MAG model. In the MAG model we make 2 main assumptions. The first MAG assumption is that all three planets in the gap are in a stable 1:2:4 MMR

$$a_2=2^{2/3}a_1; \quad a_3=2^{2/3}a_2 \qquad (2)$$

This is the most natural stable configuration for the system to evolve into as the planets all migrate inwards. Mechanisms of MMRs formation are widely studied as the result of planetary migration (see a review by Papaloizou & Terquem 2006, and references therein). As soon as this migration starts the planets open their individual gaps and then produce one large gap since their individual $4R_H$ clearing zones overlap. At this point the inwards migration nearly stops, and the orbits are stable and "locked" in the MMR (see for example figure 3 of Gozdziewski & Migaszewski 2014). The second MAG assumption is that the mass of the inner planet is twice that of the middle planet which, in turn is twice that of the outer planet ($M_{p1}=2M_{p2}=4M_{p3}$) to be compatible with current observations as noted above.

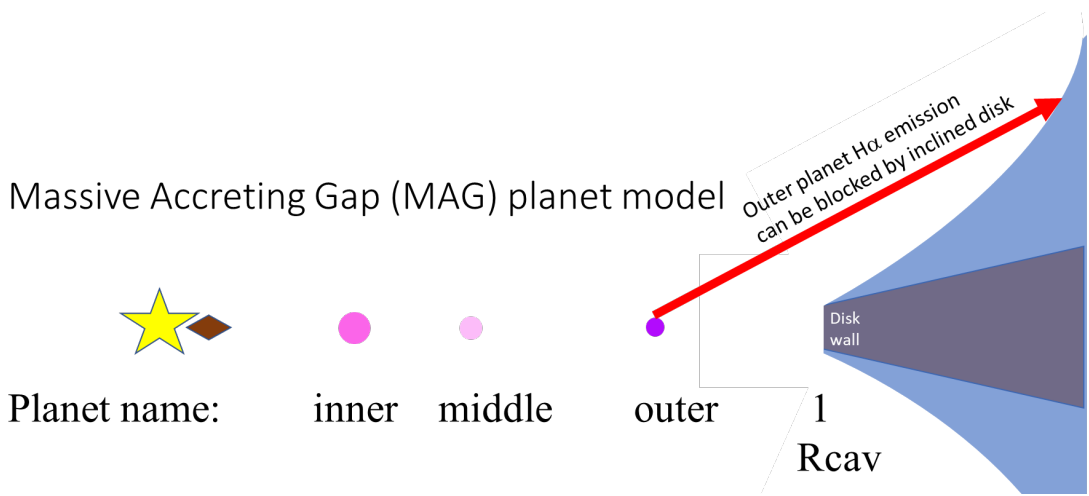

**Fig 3:** A cartoon of the MAG model of gap planets. This model satisfies the following: 1) overlapping ≤$4R_H$ sectors to be dust-free to Rcav; 2) dynamically stable: 1:2:4 MMR; 3) Each planet, pair-wise, has a separation of at least $2\sqrt{3}$ mutual Hill Radii ($R_{H,m}$); to be stable>>10Myr; 4) outer planet $4R_{H3}$ from Rcav to create dust-free edge. Planet size scales with mass.

To keep the gap edge (distance 1 Rcav from star) dust free we know that there must be 4 $R_{H3}$ between that edge and location of the outer planet $a_3$. So,

$$\text{Rcav}-4R_{H3} = a_3, \quad \text{or} \quad R_{H3}=(\text{Rcav}-a_3)/4 \qquad (3)$$

Now for planet-planet stability we require there to be 4.5 Hill spheres between the middle and outer planet (the mutual hill spheres are also calculated for all the planets in Table 1). Hence, if we need there to be 4.5 $R_{H3}$ between the middle and outer planet for stability then from equations 2 and 3 we can write:

planet-planet distance = $a_3 - a_2 = a_3(1 - 2^{-2/3}) = 4.5 R_{H3} = 4.5(R_{cav} - a_3)/4$

solving the above for $a_3$ yields:

$$a_3 = \frac{R_{cav}}{1 + \frac{4}{4.5}\left[1 - 2^{-\frac{2}{3}}\right]} = 0.753 R_{cav} \qquad (4)$$

for all cavities in the MAG model. Applying equation 4 to equation 2, yields that $a_2 = 0.475 R_{cav}$, and $a_1 = 0.300 R_{cav}$ in the MMR.

There is, of course, a range of possible stable triple gap planet solutions and we can explore their upper and lower bounds. The minimum value of $a_3$ for a stable solution is when the planet-planet distance is less than 4 $R_{H3}$. So the case that $a_{3min} - a_2 = 4R_{H3}$ then equation 4 yields $a_{3min} = 0.730$ $R_{cav}$. On the other hand, the maximum $a_{3max}$ is at the point that the planet-planet distance is $>8R_{H3}$ at which point dust is no longer scattered away between the outer and middle planet (in this case the large gap would disappear and instead appear as a pair of narrow dust free "rings" in ALMA imaging). From equation 4 we find if $a_{3max} - a_2 = 8R_{H3}$ then $a_{3max} = 0.843 R_{cav}$. In summary, we see the possible systematic error is $a_3 = 0.753^{+0.09}_{-0.02} R_{cav}$, $a_2 = 0.470^{+0.06}_{-0.01} R_{cav}$, and $a_1 = 0.300^{+0.03}_{-0.01} R_{cav}$. So the true semi-major axis could be as much as +12% to -3% w.r.t. the MAG values of $a_3$, $a_2$, and $a_1$. However, these errors are at the extrema and so the likely error is well within 10% which is an acceptable level of accuracy for the MAG model.

Since the outer planet is $4R_{H3}$ from the edge of the gap ($R_{cav}$), there must a certain mass for the outer planet mass ($M_{p3}$) to achieve this, we see from equation 3 and 1 that:

$R_{H3} = (R_{cav} - a_3)/4 = (33/40)(a_3/10)(M_{p3}/M_*)^{1/3}$ au,

but substituting equation 4 ($a_3 = 0.7525 R_{cav}$) and solving for $M_{p3}/M_*$ yields:

$$\frac{M_{p3}}{M_*} = \left[\frac{100(1 - 0.7525)}{33(0.7525)}\right]^3 = 0.99 \qquad M_{p3} \text{ in units of } M_{jup} \qquad (5)$$

hence equation 5 can be re-written as:

$M_{p3} = 0.99(M_*/M_{sun})$ in units of $M_{jup}$

So, the more massive the star ($M_*$) the more massive the planets ($M_p$). See Fig 3 for a cartoon of the MAG model, and Table 1 for a summary of the model's parameters. It is important to note, that the MAG model has no free parameters w.r.t. the orbits: the outer planet separation is simply given by:

$$sep_3 = \frac{0.753 Rcav}{D}\left[1 + \frac{2-\pi}{\pi}(1 - \cos(disk_{inclination}))\right] \quad \text{arcsec} \quad (6)$$

where $sep_3$ is the average projected separation on-sky of the outer planet. We cannot, of course, know precisely the orbital phase (or position angle PA) of the planet, just an average projection onto the sky is all that can be predicted. But since the planets are co-orbital and co-aligned with the disk cavity, the projected orbits trace the cavity as projected on the sky. We can illustrate this by superimposing the predicted orbits from the MAG model onto the PDS 70 b and c system in Fig 2. We note that MAG's orbital calculations are consistent with the orbits of PDS70b and c (see Fig 2). For example, MAG predicts from PDS 70's 76 AU gap that the inner planet ("b") should be at 0.3*76 = 22.8 au and the middle planet "c" at 0.475*76=36.1 au. Indeed, PDS 70 b is observed to be at $20^{+3}_{-4}$ au and PDS 70 c is observed to be at $34^{+12}_{-6}$ au (Wang 2020); both are highly consistent with the MAG model calculated values. This gives a useful reality check of the MAG orbital predictions.

**Table 1:** Orbits and Masses of MAG Planet Model

|  | Inner Planet #1 | Middle Planet #2 | Outer Planet #3 |
|---|---|---|---|
| **Orbital semimajor axis (*a*) as fraction of disk gap (Rcav)** | 0.300 Rcav | 0.475 Rcav | 0.753 Rcav |
| *a₂-a₁* **in Mutual Hill Spheres** | \multicolumn{2}{} $a_2-a_1$=4.5 $R_{H,m12}$ |  |
| *a₃-a₂* **in Mutual Hill Spheres** |  | \multicolumn{2}{} $a_3-a_2$=5.6 $R_{H,m23}$ |
| **Periods (1:2:4 MMR)** | $P_1$ | $2P_1$ | $4P_1$ |
| **Planet mass ($M_p$) in $M_{jup}$ (star mass ($M_*$) in Solar units)** | 3.96$M_*$ | 1.98$M_*$ | 0.99$M_*$ |

### 3.2. Hα Emission and Extinction in the MAG Model

How does the MAG model estimate the expected amount of Hα emission from a planet? The gas will "seek out" any gravitational potential wells (planets) on its slowly decaying orbit around the star – and these wells (planets) will emit in Hα as a fraction of the gas magnetospherically accretes from the circumplanetary disk onto the high polar regions of the planets. A detailed physical model of this explains PDS70b's Hα emission (Thanathibodee et al. 2019). This model predicts a mostly dust-free line of sight and significant Hα observed for the small (~2-4$M_{jup}$) planets like PDS 70c, b in agreement with the 3D thermo-hydrodynamical models of Szulagyi & Ercolano 2020, but only in their "gas only" case --this case best matches the observed properties of PDS 70b, c.

Why are only "dust free" models applicable to PDS 70b and c? If any dust falls into the gap it is scattered by the outer planet, leaving the middle and inner planets almost completely dust-free. This strongly suggests that in the case of the inner and middle planets they should best fit by the "gas only" models of Szulagyi & Ercolano (2020). Which explains why the Hα from PDS 70 b and c are detectable even though the masses of these planets are low.

### 3.3. Hα Line Luminosity Calculation by MAG model: Example of PDS 70 b

As is already published in section 3.4 of Close et al. (2014) the $L_{H\alpha}$ luminosity can be calculated for a gap planet by comparing the flux with Vega in the usual manner:

$$L_{H\alpha} = 4\pi D^2 vega_{zero_{point}} \Delta\lambda 10^{\frac{Rmag-p-H\alpha}{-2.5}}$$

where Rmag_p_Hα is just the effective de-extincted "R magnitude" w.r.t. Vega for planet "b" at Hα (this is good approximation since the center of the R filter (658nm) is almost exactly that of Hα (656.3nm)). It is clear that Rmag_p_Hα is equal to just the observed R mag of the star ($R_A$) minus the extinction to both the star and the planet ($A_R$). There is also the possibility that there is extra extinction towards the planet ($A_p$) in addition to $A_R$. There is also a slight correction for the leakage of the primary's continuum into A's Hα measurement, which causes ΔmagHα to be slightly larger than it should be. This an easy photometric correction to make by the subtraction the ratio of the aperture fluxes in the 2 SDI filters converted to magnitudes. Typically, the aperture flux of the primary in the Hα filter is ~1.3x (-0.3 mag) that of the continuum (since the accreting star is also brighter at Hα then the continuum). There is also a very slight correction in the other direction since there is extra ~0.05mag added due to Hα light in R filter. So, the "R mag" of the planet is:

$$\text{Rmag\_p\_H}\alpha = (R_A - A_R) + (\Delta \text{magH}\alpha - 0.3 + 0.05) - A_p$$

In the case of PDS 70b we have very little extinction to the star ($A_R \leq 0.2$ mag) and we assume $A_p$ is zero (Wagner et al. 2018; Thanathibodee et al. 2019). Therefore, the line luminosity $L_{H\alpha}$ can be written:

$$L_{H\alpha} = 4\pi D^2 Vega_{zero_{point}} \Delta\lambda 10^{\frac{R_A - A_R + \Delta magH\alpha - 0.25 - A_p}{-2.5}} \quad (7)$$

Which we can directly solve for in the case of PDS 70 b as:

$\text{Log}(L_{H\alpha}/L_{sun}) = \log(4\pi(113*3.1\times10^{18})^2 * 2.34\times10^{-5} * 0.006 / [3.9\times10^{33} * 10^{((11.696-0.2+(7.36\pm0.47)-0.25)/2.5)}]$

$= -5.70 \pm 0.19$ where the Vega zero point magnitude in our Hα filter ($Vega_{zeropoint}$; see Males et al. 2014 for the zeropoint calibration of our $\Delta\lambda=6$nm wide Hα filter; or spectral resolution R=109) is calculated to be $2.339\times10^{-5}$ ergs/(s cm$^2$ μm). This a significant amount of emission and almost 10x better contrast than at H band (Keppler et al. 2018). We note that PDS 70c is almost undetectable at H band (Mesa et al. 2019), but quite detectable at Hα (Haffert et al. 2019) where it was discovered. Therefore, an SDI survey in Hα should be very sensitive to very low-mass gap planets that might be very difficult to detect at any other wavelengths. The utility of Hα SDI to detect otherwise very faint gap planets was first predicted by Close et al. (2014).

Since low mass, young, objects have excellent Xshooter calibrated accretions rates (Rigliaco et al. 2012) we can use:

$$L_{acc} = 10^{[2.99\pm0.23 + (1.49\pm0.07)*(\log(LH\alpha))]}$$

from the empirical total accretion luminosity $L_{acc}$ to $L_{H\alpha}$ relations of Rigliaco et al. (2012) for very low mass accretors.

However, Thanathibodee et al. (2019) find, by applying the first full treatment of Hα line radiative transfer in a magnetospheric geometry for planetary-mass objects, that weakly accreting planets accreting with $\dot{M}_p < 5\times10^{-12} M_{sun}$/yr (similar to PDS 70c) are better fit with a $\log(\dot{M}_p)$ varies as $0.353\log(LH\alpha)$ power-law, so the production of Hα is less efficient. Hence, we find that in this case this a power-law:

$$L_{acc} = 10^{[-3.62 + 0.353\log(LH\alpha)]}$$

better describes the relationship between Hα luminosity and total luminosity for low planetary accretion rates. Which yields a lower accretion luminosity ($L_{acc}$). For example, in the case of PDS 70b the relationship yields $\log(L_{acc}/L_{sun})=-5.63\pm0.18$. Then using the standard relation relating the released total accretion luminosity $L_{acc}$ from accretion onto the planet surface:

$$\dot{M}_p = 1.25 L_{acc} R_p/(GM_p)$$

of Gullbring et al. (1998), yields a planetary accretion rate of $\dot{M}_p=5\times10^{-12}$ $M_{sun}$/yr (using $M_p$ mass estimate of ~4 $M_{jup}$ for PDS 70b; Wang et al. 2020). Planet radii ($R_p$) are from 5 Myr COND evolutionary model estimate of $R_p=1.3R_{jup}$ (Baraffe et al. 2003).

*3.3.1. What Fraction of the Stellar Accretion Rate Accretes onto a Gap Planet in the Model?*

This planetary accretion rate for PDS 70b is $5\times10^{-3}$ $M_{jup}$/Myr which suggests the ~5Myr planet is at the end of its accretion phase. But even in this rather weak accretion flow, the Hα emission allowed both PDS 70 b and c to be detected. This planetary accretion rate is also equal to ~10% of PDS70 A's stellar mass accretion rate $\dot{M}_*\sim6\times10^{-11}$ $M_{sun}$/yr (Thanathibodee et al. 2020). Therefore, we will adopt *f*=10% as an estimate of the fraction (*f*) of the stellar accretion that is captured by a gap planet in the MAG model. For simplicity, we will assume this value holds for all MAG gap planets as a global average rate, although we are cognizant that this rate could vary between different planets and systems, but we hope that applied over large number of systems will be helpful as guide. Indeed, Lubow et al (1999) predicts the planetary gas capture rate could be as high as 50%, yet in the case of 3 planets this would effectively cut off accretion to the star itself (leaving just <12%) and would also imply rapid planet growth to brown dwarf masses. Since stellar accretion is observed for all our disk sample, and we removed any brown dwarf or stellar companions (eliminating binaries HD142527AB, GG Tau Ab, V4046 Sgr AB), we will adopt that 10% of the stellar accretion is captured by each of the 3 gap planets (*f*=10%) in the MAG model.

Hashimoto et al. (2020) could not significantly detect PDS 70b at Hβ with MUSE and claim $A_p\sim2$ mag since Hβ/Hα<0.3. They find $\dot{M}_p \geq 5\times10^{-11}$ $M_{sun}$/yr which is larger than the stellar $\dot{M}_*$ from A, suggesting *f*>100%. This very high *f* value is questionable since Hβ/Hα could be ~0.3 for planets (with Av=0) and also Hβ is very hard to measure as the Strehl at 486.1nm is <4% so b's weak Hβ is smeared out. Hence, we continue adopting *f*=10% for the survey as a whole and for PDS 70 b and c.

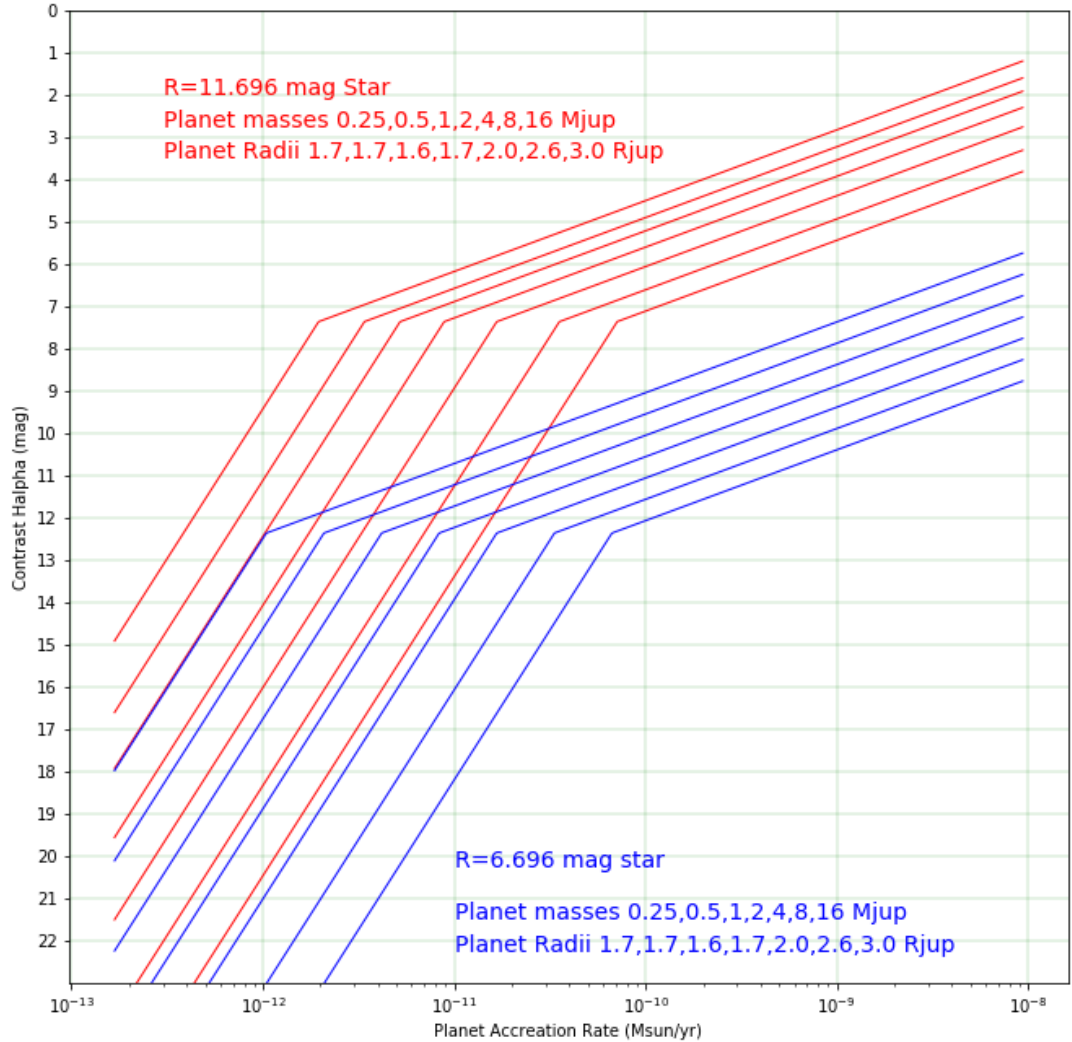

Fig. 4: Plots of estimated ΔmagHα vs. $\dot{M}_p$ from equations 8 and 9. Note the sharp drop in Hα emission as $\dot{M}_p < \dot{M}_{p\_cross}$ (as the mass accretion rate onto the planet falls below the crossing point and magnetospherical accretion is less efficient at generating Hα). We plot the contrasts from a 1 Myr old "faint" (R=11.696 mag; like PDS 70 A) star on the top. Red lines are planet masses from 0.25 $M_{jup}$ (lowest line) to 16 $M_{jup}$ (highest line). All planet radii are from COND models. We also show the strong 100x increase in contrast required to detect the same population of planets around an otherwise identical "bright" (R = 6.696 mag) star. We illustrate here that the brightest, popular, AO stars can be poor (higher-contrast) targets for direct detection of Hα planets.

## 4.0 MODEL PREDICTIONS

*4.1. Model Predictions of the Hα Contrast of Gap Planets: Cases of High and Low Accretion*

The MAG model can predict the planet/star contrast (Δmag at Hα) from the work above. For each planet there is a crossing point in the amount of the accretion flow onto that planet ($\dot{M}_{P\_cross}$) where accretion passes from being efficient to less efficient. In the high accretion case ($\dot{M}_* > 5 \times 10^{-11} M_{sun}/yr$) we can show from equation 7 and the $L_{acc} = 10^{[2.99 \pm 0.23 + (1.49 \pm 0.07)*(\log(L_{H\alpha}))]}$ relations of Rigliaco et al. (2012) that (note all terms are in standard astronomical cgs units, except D which is in pc):

$$\Delta mag H\alpha = -1.675[\log M_p + \log \dot{M}_p - \log R_p] + A_R + A_p - R_A + 5\log D + 67.9 , \qquad (8)$$

but if weak accretion ($\dot{M}_* \leq 5 \times 10^{-11} M_{sun}/yr$) then $L_{acc} = 10^{[-3.62 + 0.353\log(L_{H\alpha})]}$ ; so

$$\Delta mag H\alpha = -7.08[\log M_p + \log \dot{M}_p - \log R_p] + A_R + A_p - R_A + 5\log D + 258.22 , \qquad (9)$$

Where: $M_p$ is the mass of the planet (from MAG model), $\dot{M}_p = f\dot{M}_*$ ($\dot{M}_*$ from FVDM); planet radius $R_p$ from COND models; extinction towards the star $A_R$ from FVDM; the R band ($R_A$) magnitude of the star from SIMBAD, and the distance D (in pc) to the star from FVDM. All these observational inputs to equations 8 and 9 are listed in Table 2.

The exact mass accretion rate at the crossing point when $\log(\dot{M}_p) = \log(\dot{M}_{P\_cross})$ from the high luminosity (equ. 8) to low luminosity (equ. 9) case is linearly dependent on the planet radius and inversely dependent on planet mass. If we rewrite the total accretion relation as

$$L_{acc} = 10^{[b + (c)\log(L_{H\alpha})]}$$

then we can derive an exact solution for the cross-over point from high accretion $L_{acc}(b,c)$ to low accretion $L_{acc}(b',c')$ as (in cgs units):

$$\log(\dot{M}_{P\_cross}) = \log(R_p) - \log(M_p) + (b'c - bc')/(c - c') + 40.86 \qquad (10)$$

Since c=1.49 and b=2.99 in the high efficiency case (Rigliaco et al. 2012); whereas in the lower efficiency case: c'=0.353, b'=-3.62 (adapted from Thanathibodee et al. 2019). So, from equation 10:

$$\log(\dot{M}_{P\_cross}) = \log(R_p) - \log(M_p) + 35.19 \qquad (11)$$

hence as $\dot{M}_p$ falls below $\dot{M}_{P\_cross}$ there is a power law break and decrease in the Hα flux.

Hence, we see from equation 11 (and Fig 4) that higher mass planets can remain efficient at producing Hα ($\dot{M}_p > \dot{M}_{P\_cross}$) compared to lower mass planets since $\dot{M}_{P\_cross}$ varies inversely with mass. For example, the first star in Table 2 is HD100453 for which the MAG model predicts a 5.66 $M_{jup}$ inner planet. For that inner planet $\log(\dot{M}_{P\_cross1})$=-11.68 (equation 11) which is just below the MAG estimated accretion rate of $\log(\dot{M}_{P1})$= -11.5 ($\dot{M}_p$ estimated from 10% of the observed $\log(\dot{M}_*)$=-10.5 for this star; col 2 Table 2), so since $\dot{M}_{p1} > \dot{M}_{P\_cross1}$ we can safely use equation 8 to calculate ΔmagHα for HD100453's inner planet. However, the outer planet is predicted to be 4x lower mass and as a result $\log(\dot{M}_{p\_cross3})$=-11.08. Since the $\log(\dot{M}_{p3})$=-11.5 (again just 10% of $\dot{M}_*$) then we find $\dot{M}_{p3} < \dot{M}_{P\_cross3}$ and so we are in the lower luminosity case and we need to use equation 9 to calculate ΔmagHα for HD100453's outer planet. Equation 11 allows, for every MAG planet in Table 3, its ΔmagHα to be correctly calculated by equation 8, or if $\dot{M}_p < \dot{M}_{P\_cross}$ by equation 9. In Table 3 we present all the calculated gap planet properties calculated from the observed star/disk properties in Table 2. In Table 3 the values in bold text are cases where $\dot{M}_p < \dot{M}_{P\_cross}$. As is clear from Table 3, usually $\dot{M}_p > \dot{M}_{P\_cross}$ since most of the transitional disks are accreting at much higher rates ("active accretors") than, for example, either PDS 70 or HD 100453 which are on the low end ("weak accretors").

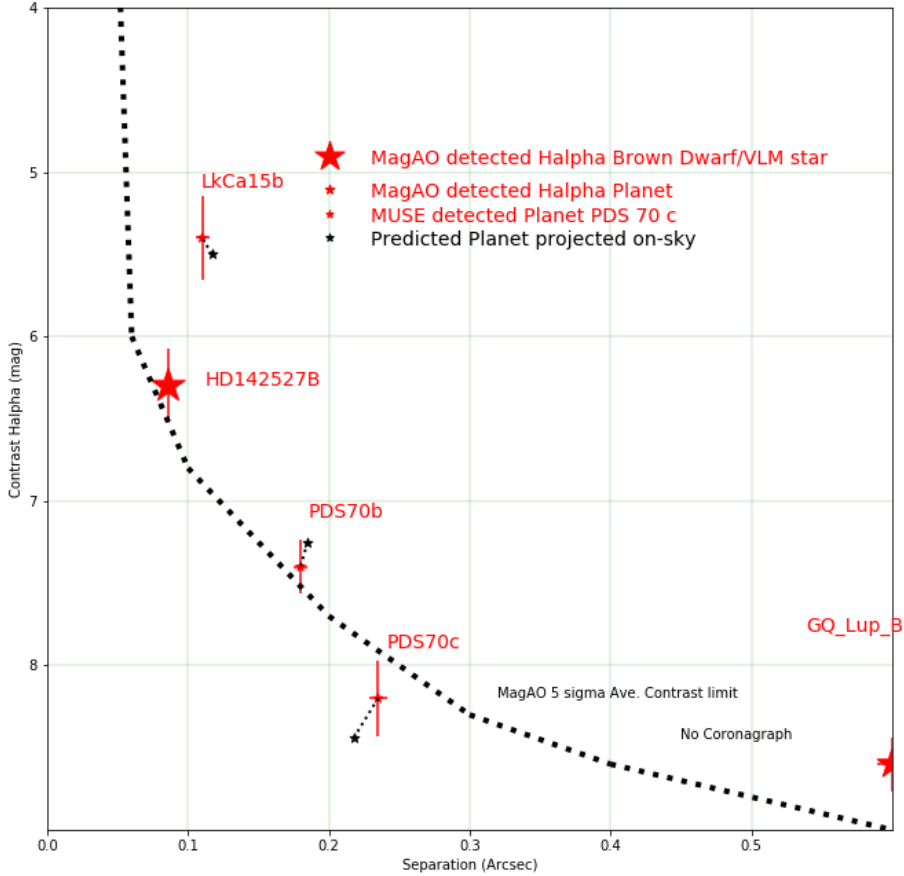

**Fig 5:** Fits of the MAG model to the previously detected Hα gap planets. Note how well the detectable planets (*small* red stars with errorbars) are predicted by the MAG model (black stars) where the black star is corrected to the observed PA (orbital phase) of the planet it represents. Stellar companion HD142527B (Close et al. 2014) and brown dwarf GQ Lup B (Wu et al. 2017a) are also shown but are not fit by MAG. GQ Lup B is actually at 0.8" but is shown here at edge of plot for completeness.

*4.1.1. The Special Cases of the Detected Hα Planets PDS 70 b, PDS 70 c, and LkCa 15 b*

We can apply the MAG model to the known Hα planets. For LkCa 15 b we take the observed planet/star contrast (ΔmagHα) from the MagAO observations from Sallum et al. (2015). For PDS 70 we take the observed ΔmagHα from MagAO observations by Wagner et al. 2018 (MagAO observations are consistent with the filter width, and photometric zeropoint used in equation 7) and the observed different between ΔmagHα of b and c from Haffert et al. (2019) from the MUSE IFU (since c has only been detected by the MUSE IFU at Hα). The results are presented in Fig 5. As can be seen from Fig 5 the model predicts Hα contrasts (ΔmagHα) consistent with observational errors. This is an encouraging sign that the MAG model can reproduce all currently known gap planet Hα contrasts and orbits. The orbital fits are quite good, particularly in the case of PDS 70 b and c. In Fig 2 we overlay the predicted MAG orbits over the observed Hα planet positions, the agreement is very good. This success encourages us to apply the MAG model to all 9 of the previously observed large-gap disks that overlap between FVDM and AO observations of Zurlo et al. (2020), plotted in Fig 6.

*4.1.2. The MAG Model Applied to the 9 Most Commonly Observed Disks*

It is straightforward to take the 9 best (age <5Myr; Rcav >20au; highest $\dot{M}_*$) and most often AO observed Hα stars and predict which gap planets should have been detected and which are still too faint/close-in to be detectable with today's AO systems at Hα. This is done in Fig 6 where ΔmagHα vs. *sep* is plotted for all stars in common between FVDM and Zurlo et al. (2019).

The most notable point to Fig 6 is that the top systems that are predicted to be most easily detected (PDS 70 and LkCa 15) have, in fact, *been* detected to have Hα gap planets. These planets are: PDS 70 b, c (Haffert et al. 2019); and LkCa 15b Sallum et al. (2015). This gives us some confidence in a reality check of the MAG model.

However, LkCa 15 c and d are predicted to be also detectable by the model and indeed multi-epoch (5-7 years) detections of point sources on Keplerian orbiting sources in Sallum et al. 2015 and 2016 have been reported as c and d. However, neither of these candidates have been clearly detected at Hα. Recently LkCa 15 "c" and "d" have been suggested to be inner disk features (Currie et al. 2019) and so they may be compact dust and not gap planets and to be conservative we will not consider these planets as unambiguously detected. But we note LkCa 15 b is a 6σ MagAO Hα source and hence a true gap protoplanet, and not a disk feature (Sallum et al. 2015) therefore it is used in this analysis. The true nature of the LkCa 15 c and d "objects" will hopefully be revealed by better AO correction in the near future. It is important to note that LkCa 15 at R~12mag and DEC +24º is a very challenging AO target for today's Hα imagers in the southern hemisphere. The lack of a middle or outer Hα planet detection suggests that innermost Hα planet might, in general, be easier to detect (the dust is scattered away by the outer planets which, in turn, extinguishes the Hα from these planets).

Also, the MAG predicted outer planet "PDS 70 d" at ~0.43" has not been yet been definitively detected --even though it might have been just possible with MUSE. Keppler et al. (2019) have recently detected with ALMA a non-Keplerian (i.e. planet-like not primary disk-like; Perez et al. 2014) $^{12}$CO point source at 6$\sigma$. At 0.39" it is near the 0.43" MAG predicted position of PDS 70 d, moreover, at PA=260º the direct line of sight. to "d" is blocked/extincted by the disk's 50º inclination and explains why it is not a detected H$\alpha$ source by MUSE in Fig 2. Further work with ALMA will be required to confirm if this "d" object is real.

*4.1.3. Extra Extinction to the Outer Planet and Cases of Dusty Gaps*

In general, H$\alpha$ from the outer planet can be blocked by an inclined disk (see red arrow in Fig 3) or is not easily detected against the glare of the disk "back wall" dust clumps. The inner 2 planets are less likely to be blocked by the flared disk, and they are much less likely to have any additional extinction as they are in the dust-free part of the disk gap, so the "gas only" accretion case applies (Szulagyi & Ercolano 2020) leading to $A_{p1} \sim A_{p2} \sim 0$ mag. Hence, outer planets can be the hardest gap planets to detect in H$\alpha$ compared to the middle or inner planets (despite having a larger separation on-sky). Hence $A_{p3}$ can be large; whereas $A_{p1}$ and $A_{p2}$ are likely close to zero. However, we don't know which outer planets are blocked by their flared disks (or dusty accretion flows Szulagyi & Ercolano 2020), so we assume $A_{p3}$ is zero. However, the gaps of HD100546, HD135344B, MWC 758, and AB Aur are famous dusty spirals and so to account for this extra gap dust, we adopt $A_P \sim 2A_R$ for each planet in these systems ($A_P$~0.8-2.4 mag) --otherwise $A_P$ is zero in Table 3.

Another interesting detail of Fig 6 is that MAG predicts only PDS 70 and LkCa 15 planets are wide and low enough contrast to be detected (purple circles above dotted sensitivity curves). The other 7 systems' planets are just a bit too tight/high-contrast to be detectable with today's H$\alpha$ imagers. Since this is exactly what we observe, we can conclude that the MAG model has some predictive power. To discover more gap planets, we clearly need higher H$\alpha$ contrasts at smaller separations.

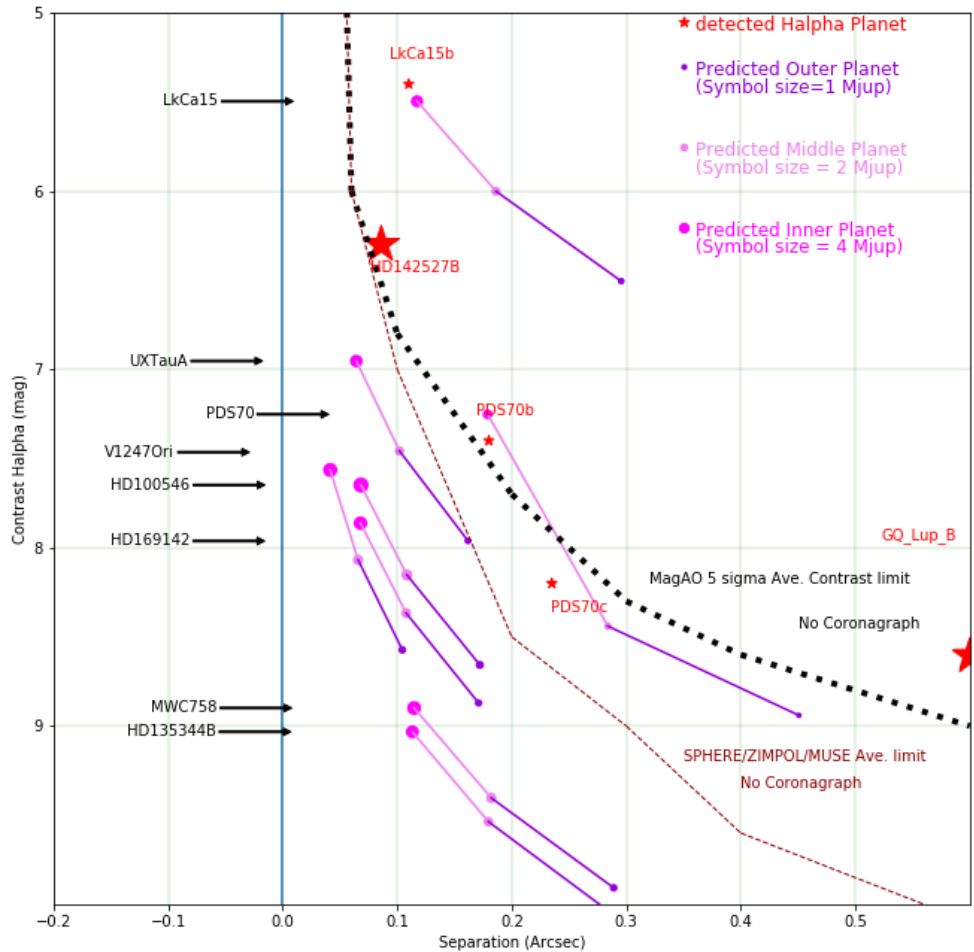

**Fig 6:** The MAG model applied to observed disks. LkCa 15 and PDS 70 appear as the best systems for Hα detections. All the disks above have been observed by ZIMPOL but no planets have been detected to date by these ZIMPOL observations (Cugno et al. 2019; Zurlo et al. 2020). To be a detectable planet at 5σ a MAG gap planet (purple circles) must be above the dashed sensitivity curves. Current AO surveys are not quite sensitive enough to detect new Hα planets (other than in the PDS 70 and LkCa 15 systems). The ZIMPOL/MUSE average curve is from a 10[th] mag star (TW Hya) and similar to 8[th] mag stars HD 100546 and MWC 758 (Cugno et al. 2019) although this average curve might be slightly too sensitive for fainter targets since ZIMPOL has yet to detect PDS 70 b, which is acceptable since this is just meant as an average sensitivity curve from ~8<R<12 mag.

*3.2. Model Predictions for the Whole Population of Actively Accreting Large Gap Disks*

We can also apply the MAG model to all 23 bright (R<13 mag) single stars from FVDM. It is likley that there are many possible gap planets that are possible to detect with slightly better contrast sensitivity at Hα. So, we need to understand what realistic contrasts might be like with advanced AO systems.

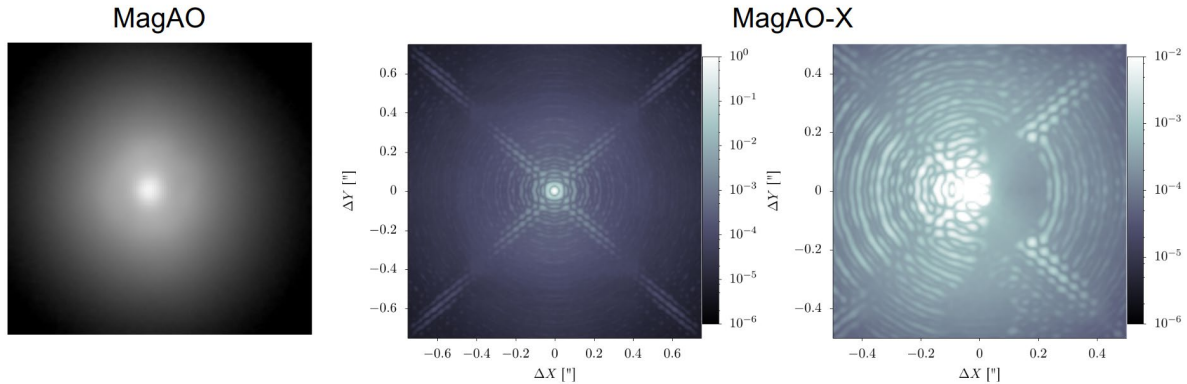

**Fig. 7:** Improvements of Hα contrasts. MagAO (Strehl ~10%; no coronagraph) on an 11$^{th}$ mag star vs. simulated MagAO-X PSF on a 10$^{th}$ mag star in good seeing (50% Strehl). MagAO-X vAPP coronagraphic PSF to the far right. The raw contrasts are predicted to be ~10$^{-3}$ at 100 mas (4.57λ/D) with MagAO-X. After 1hr and ASDI KLIP pipeline reductions (Males et al. 2014a) the contrasts will approach 10$^{-4}$ at 100 mas, and then 10$^{-5}$ contrasts at 150 mas inside the "dark hole" (area of highest contrast to right of star) created by the interference effects of the vAPP (see solid blue line in Fig 8). This is a factor of >100x better contrast than what MagAO could achieve. Colors are raw PSF contrast.

*4.2.1. Contrast Curves for Future Surveys with MagAO-X*

As we increase the number of controlled modes (~1500), increase the speed >2kHz, while simultaneously removing the readnoise by use of EMCCD wavefront sensors, the AO community is achieving higher and higher Strehls at Hα. MagAO-X with its impressive ~15cm/actuator mapping of the telescope pupil has been designed from the start for Hα planet imaging (Males et al. 2018; Close et al. 2018). MagAO-X has a successful first light in December 2019. Results are reported in Males et al. (2020). Here we will use the Hα contrast curves for MagAO-X from Males et al. (2019). These contrasts are based from full "end-to-end" models of the MagAO-X system. In Fig 7 we show a comparison of how PSF quality and contrasts at Hα can improve with MagAO-X. From these PSFs and full modeling of long 1hr datasets with correct treatment of photon-noise and residual speckle noise, we present the 1hr contrasts of MagAO-X as the solid blue curve in Fig. 8. This curve is consistent with the initial first light results of Males et al. (2020).

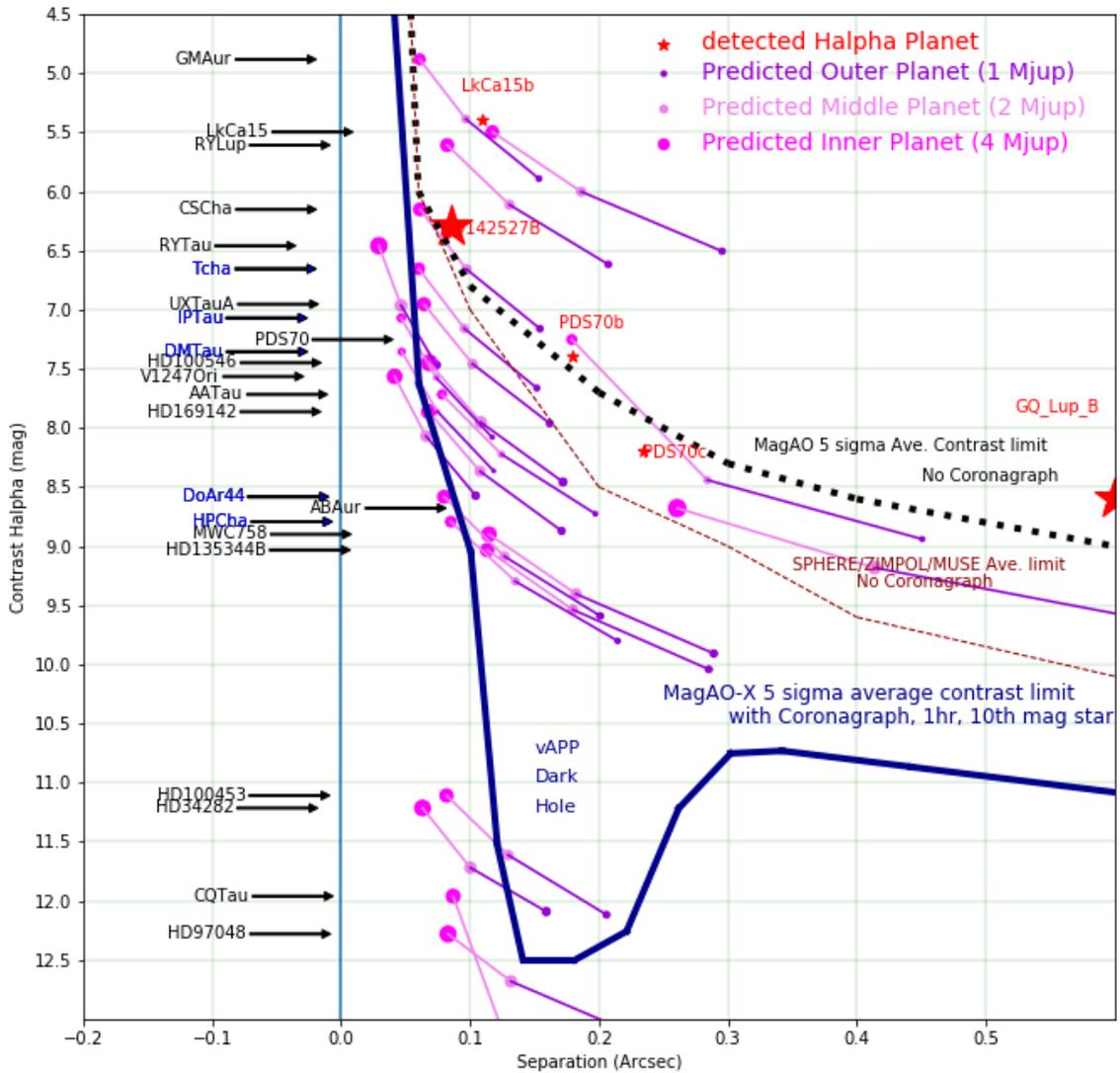

**Fig. 8:** The MAG planet model applied to all 23 bright (I<13) FVDM single stars. The MagAO-X 1 hr. contrast limits from processing the images in Fig 6 are plotted as the solid blue line. We can trivially see that MagAO-X should be able to detect ~48 gap planets (44 new) from the top 19 systems (all those above HD100453).

*4.2.2. How to Weight Contrasts for Faint and Bright AO Targets for Same Plot*

The 5 faintest targets are labeled in blue colored text in Fig 8 (12<$R_A$<13 mag) and they each have the contrasts increased by 2.0 mag to account for the lower AO Strehls for these stars. These faint stars could be executed in better than median observing conditions (good, slow, seeing), hence the blue curve is a good guide for all targets (6<$R_A$<13 mag) even if the contrasts will be ~2.0 mag worse at $R_A$~13 mag --because each faint "blue" target has had +2.0 mag added to the contrast in Fig 8 to compensate for this effect. This increase in contrast is reasonable because contrast scales as Strehl/(1-Strehl) and the Strehl is very low for these faint R>12 targets. From space (or with an excellent laser guide star system) these targets would be quite promising since the Strehl loss would not apply and these systems would have 2 mag smaller contrast than shown in Fig 8.

## 5.0 DISSCUSSION

### 5.1. Some Famous Systems Not Included

*5.1.1. Rings vs. Wide Gaps in Transitional Disks: The Case of TW Hya*

Fig 8 clearly shows that many disks look like excellent targets for future H$\alpha$ observations. Right at the top (in terms of favorable $\Delta$magH$\alpha$) would be a star we did not include: TW Hya. TW Hya would be oldest object in this list at ~10 Myr. But it would also be the closest at 60 pc, which gives the self-luminous H$\alpha$ from the planet a boost, and it is almost face-on. However, ALMA imaging of TW Hya doesn't show a very wide gap. It would be better described as cleared narrow "ring". In case of TW Hya the ring is on the order of just ~2 au wide (FVDM) so it must be cleared by a single gap planet at *a*~20 au (so the 3 planet MAG model doesn't apply). A single "ring" planet may suffer dust self-extinction making detection at H$\alpha$ problematic (Szulagyi & Ercolano 2020) hence TW Hya "b" may prove to be a difficult planet to detect at H$\alpha$. Such rings are quite common in ALMA high-resolution images, like those of DSHARP (Andrews et al. 2018). For example, the famous ALMA images of HL Tau show separate gaps or rings where 3 near-resonance massive planets can explain the observed gaps and have stable orbits >10Gyr (Wang et al. 2020b) but those planets are each located in their own gap, and the N-body simulations of Wang et al. (2020b) do not consider all 3 planets in the same gap, yet they still argue that systems of 3 near-resonance massive planets can be stable in a gapped disk. However, a single massive planet can sometimes clear multiple rings (Zhang et al. 2018). But these rings are usually quite narrow, unlike the 20-80 au wide dust free gaps that we list in Table 2 which each need multiple planets (Zhang et al. 2018). Hence, "ring" transitional disk systems are inappropriate for the MAG model. The MAG model only works for transitional disks with large (>20au) dust free gaps.

*5.1.2. Circumbinary Rings vs. Wide Gaps: HD 142527, GG Tau, and V4046 Sgr*

We have tried to not include binaries because they do not require gap planets to clear their gaps since a single stellar companion will suffice (such as HD142527B; Close et al. 2014 or GG Tau Ab; Roddier et al. 1996). In the case of V4046 SgrAB the binary is very tight and so may not be responsible for clearing the gap, and gap planets could be possibly present. However, the gap in V4046 is still a circumbinary disk and so is not appropriate for the MAG model.

*5.2. Inclined Systems*

In Fig 8 we can see some systems that look quite promising include RY Lup, RY Tau, and T Cha. All these systems have favorable predicted $\Delta$magH$\alpha$ but we strongly caution here that they also have quite inclined disks (FVDM). Hence, any potential gap planets in these inclined systems may prove very difficult to detect at H$\alpha$ due to dust extinction from their flared disks along the line of sight.

*5.3. Low Accretors*

Four disks (HD100453, CQTau, HD34282, and HD97048) did not have a single planet with $\Delta$magH$\alpha$ < 11 mag. These stars have only upper limits (<$10^{-11}$ $M_{sun}$/yr) to the stellar mass accretion rate. Hence, there is not enough gas flow to produce a detectable amount of H$\alpha$ from the planetary accretion. They were also hurt by the fact that their primaries were quite bright at (R~7-9$^{th}$ mag), clearly a gas-starved planet around a bright star is a poor H$\alpha$ target, even if it is a good (bright) AO guide star (as shown by Fig 4).

This highlights another often overlooked bias in AO surveys where the brightest stars tend to be executed first (and best) since the AO correction is better and the observations are easier. However, as demonstrated in Fig 4 many of these bright stars need even higher contrasts to detect a planet --especially if there is little gas in the system. Hence, lower luminosity targets like PDS 70 and LkCa 15 (both with $R_A$~11.7 mag) have relatively small $\Delta$magH$\alpha$. These disks particularly become excellent targets when their very large (76, 72 au; respectively) gap sizes are also considered. Hence PDS 70 and LkCa 15 are the very best wide gap planet targets in the sky.

### 5.4. The Southern Bias

It is also clear from comparing Fig 6 to Fig 8 that there are many excellent targets that have not been observed yet at H$\alpha$ with MagAO or VLT/SPHERE/ZIMPOL or MUSE. Some examples include AB Aur, GM Aur, and RY Tau. The reason that these have been not as closely studied at H$\alpha$ is that, other than the 8m Subaru's SCExAO system (Uyama et al. 2019), there are no high-contrast H$\alpha$ imagers in the northern hemisphere, so these DEC $>+25°$ targets are challenging for southern AO systems. But Fig 8 suggests GM Aur and AB Aur could be excellent targets and worth the effort (while RY Tau suffers from a high inclination).

### 5.5. What is the Expected Yield of a Future MagAO-X Survey?

From Fig 8 we predict with the MAG model a maximum yield of 48±7 planets from 19 stars (44 would be new discoveries) in a future MagAO-X survey. This planet yield is robust, for example a "flat" mass ratio model ($M_1=M_2=M_3$) applied to our sample would suggest a similar planet yield of ~43 planets. So, this prediction is not locked into the fine details of the MAG model alone.

In the farther future, advanced AO on the 25m GMT (like GMagAO-X; Males et al. 2019; Close et al. 2020) or the 30m TMT (like PSI; Fitzgerald et al. 2019) telescope, will be able to detect all gap planets down to ~3au orbits (Sallum et al. 2019).

### 5.6. What is the Expected Yield of a Future Survey if the Outer Planets are not Detected?

It is possible that all the outer planets prove hard to detect. This could be due to extinction by the disk edge (as could be the case for PDS 70 d), or due to higher intrinsic extinction ($A_p$) due to the outer planet receiving the bulk of the in-falling dust to scatter, or simply because there are only 2 massive (inner + middle) planets and the outer planet is much less massive. Whatever the cause, we can see from Fig 8 that we still have 25±5 new inner and middle planets detected even if outer planets are hard to detect. So, such a future survey will be productive regardless of the fine details of the true gap planet population.

### 5.7. Sensitively of the Predicted Planet Yields to Assumptions About the Planet Masses

The MAG model makes 2 main assumptions. One that the 3 planets are on an MMR spaced correctly so that the gap is cleared. This is a reasonable assumption based on stability of multi-planet wide systems like HR 8799bcde. The second is that the middle planet is 2x the mass of the outer planet and the inner planet 2x that of the middle planet ($M_{p1}=2M_{p2}=4M_{p3}$). This assumption is inspired by observations of all known wide multi-planet systems (PDS 70, HR 8799, TYC 8998-760-1 and even that of our own Solar System). However, it could easily be that the mass ratio instead of 2x is in a range of 1-3x. In the case that the mass ratio is 1.0x (all planets same mass as $M_3$) that would have no affect on $\Delta magH\alpha_3$ but $\Delta magH\alpha_2$ would be +0.5 mag worse (bigger) and $\Delta magH\alpha_1$ would be +1.0 mag worse than shown in Fig 8. On the other hand, if the mass ratio increased to 3.0x (as in our Solar System) then $M_{p2}=3M_{p3}$ and $M_{p1}=3M_{p2}$. In that case $\Delta magH\alpha_3$ would still be the same but $\Delta magH\alpha_2$ would be 0.3 mag better (smaller) and $\Delta magH\alpha_1$ would be 0.59 mag better (smaller). In other words, allowing a wide range of mass ratios would not affect the $\Delta magH\alpha_3$ values, but it could change $\Delta magH\alpha_2$ by -0.3 to +0.5 mag and $\Delta magH\alpha_1$ by -0.59 to 1.0 mag compared to those values in Fig 8. While these are significant differences they do not significantly affect the planet yields or major conclusions of this study. For example, the planet yield in the "worst case" of all the planets with same mass as $M_{p3}$ ($M_{p1}=M_{p2}=M_{p3}$) is 42 systems of the 19 best stars (quite consistent with the 48±7 found with the default $M_{p1}=2M_{p2}=4M_{p3}$). Hence, the MAG model's predicted planet yields are not highly sensitive to exact planet mass ratios. But it would be healthy to consider a possible range of -0.59 to +1.0 mag systematic error in our predictions of $\Delta magH\alpha_1$ to cover for possible mass errors of overestimating $M_1$ by 4x and also for possibly underestimating $M_1$ by 2.25x. Similarly, the systematic errors of $\Delta magH\alpha_2$ could be as high as -0.3 to +0.5 mag. But neither of these error ranges significantly changes the planet yields.

It is certainly true that there will be a range of mass ratios in nature and we do not expect the MAG model's $M_{p1}=2M_{p2}=4M_{p3}$ to hold for every system, at best it might prove to be an average relation over all systems. However, as noted above, the predicted planet yields are consistent with 48±7 from $M_{p1}=M_{p2}=M_{p3}$ to $M_{p1}=3M_{p2}=9M_{p3}$. Hence, we do not consider the exact planet mass ratio relation to have to be known, *a priori*, to be able to estimate planet yields with the MAG model.

*5.8. What Would a Null Result Mean of a Future Deep Hα Planet Survey?*

Even in the *very* unlikely case that LkCa 15b and PDS70b,c are the only gap planets detectable at Hα in nature (hence a future survey "null result"), the great sensitivity with MagAO-X to low-mass planets will allow astronomers to place tighter constraints on the outer extrasolar population than ever before possible (for both "cold-start" and "hot-start" planets; at Hα both are equally visible; Sanchis et al. 2020) --and such a "null" result would then reject the MAG and DRS multiplanet clearing gap models at ~5σ significance. A future survey null result could also suggest that planetary accretion could be quite time variable (like FU Ori; Brittain et al. (2020), or that Hα produced by planetary accretion is heavily extincted as suggested by the dusty models of Szulagyi & Ercolano (2020) and the low extinction towards PDS 70 b, c and LkCa 15 b was an outlier. However, we strongly suspect that PDS 70 b, c and LkCa 15 b are not the only Hα planets, and that future higher-contrast surveys will discover many more Hα planets (but these new planets will typically be higher-contrast, and closer-in, than the PDS 70 planets).

*5.9. Future Theoretical Work for the MAG Model*

The MAG model presented here is a very simple model that can predict the rough properties of planets clearing the gaps of wide transitional disks. While, these gap planet MMRs can be stable for millions of years (as shown by the detailed 2D HD simulations of Bae et al. 2019 in the case of PDS 70 b and c) future detailed simulations considering both orbital migration and mass growth would will help inform the full history of these gap planet's orbital evolution. For example, can they stay dynamically stable through phases of high rates of planet growth? In particular, it would be interesting to model in detail how the trapping of the planets into the MMR occurs during the slow orbital migration inwards (or outwards; Bae et al. 2019) for all the accreting planets. During that period the planets must avoid: orbital instabilities; exciting the eccentricities; and close interactions between the planets. While this has been shown to work in the past (DRS model; Bae et al. 2019; Gozdziewski & Migaszewski 2014) more long-term evolutionary HD simulations with planetary accretion would be illuminating.

## 6.0 CONCLUSIONS

Sub-mm interferometry (SMA, ALMA etc.) has detected a significant group of large (20-80 au) gaps in many transitional disks (FVDM, references within). A handful of these disks have been shown to have H$\alpha$ bright companions inside them (HD142527B; Close et al. 2014; LkCa 15 b; Sallum et al. 2015; PDS 70b; Wagner et al. 2018; PDS 70c; Haffert et al. 2019), but some transitional disk AO surveys have not revealed any new gap planets (Cugno et al. 2018; Zurlo et al. 2019). This has encouraged recent theoretical studies which suggest H$\alpha$ can be only detected from the most massive planets (>10 $M_{jup}$) Szulagyi & Ercolano (2020) or can be highly variable (Brittain et al. 2020). But are these null results a selection effect of the AO sample selected and the limits of high-contrast AO at H$\alpha$? To answer this question requires a simulated parent population of gap planets applied to all known wide gap transitional disks.

We have presented a massive accreting gap (MAG) planet model that ensures these large gaps are kept dust free by the scattering action of 3 co-planar planets in a 1:2:4 MMR. With few free parameters, our model is consistent with the observed separations and H$\alpha$ fluxes for LkCa 15 b and PDS 70 b and PDS 70 c within observational errors. Moreover, the model suggests that the scarcity of detected H$\alpha$ planets is likely a selection effect of the current contrast limitations of non-coronagraphic, low Strehl, H$\alpha$ imaging with older AO systems. We predict that, as higher Strehl AO systems (with high-performance custom coronagraphs; like MagAO-X) are utilized at H$\alpha$, the number of detected gap planets will substantially increase by, as much as, tenfold. Detections of a large number (~25-44) of new accreting protoplanets will significantly improve our understanding of planet formation, solar system architectures, and planet disk interactions.

## ACKNOWLEDGMENTS

We would like to thank the anonymous referee for many helpful suggestions leading to a much improved manuscript. We would like to thank Jared Males for allowing figure 11 from the MagAO-X PDR (Males et al. 2019) to be reproduced here and for helpful discussions. We thank Logan Francis and Nienke van der Marel for permission to reproduce their ALMA images in Fig 1. We would also like to thank Sebastiaan Haffert (for the background for Fig 2) and Alycia Weinberger both for helpful discussions about planetary accretion. We thank Kate Follette for helpful discussions about gap planets and their environments. Laird Close was supported by a NASA eXoplanet Research Program (XRP) grant 80NSSC18K0441.

*Facilities*: Magellan:Clay (MagAO); Magellan:Clay (MagAO-X)

**Table 2:** A List of the Observed Parameters of the Large Gap Transitional Disks

| Transitional Disk System Name | Log $\dot{M}_*$ ($M_{sun}$/yr) | $R_{p3}$ ($R_{jup}$) | $A_R$ (mag) | $R_A$ (mag) | D (pc) | $R_{cav}$ (au) | $M_*$ ($M_{sun}$) | $L_*$ $L_{sun}$ | *Approx. inclination* |
|---|---|---|---|---|---|---|---|---|---|
| HD 100453 | ~ -10.5 | 1.3 | 0.0 | 7.5 | 104 | 30 | 1.47 | 6.2 | 30° |
| HD 100546 | -7.04 | 1.4 | 0.4 | 6.5 | 110 | 29 | 2.13 | 25 | 50 |
| HD 135344B | -7.37 | 1.5 | 1.2 | 8.6 | 136 | 52 | 1.51 | 6.7 | 10 |
| HD 169142 | -8.7 | 1.4 | 0.4 | 8.3 | 114 | 26 | 1.65 | 8.0 | 0 |
| LkCa 15 | -8.4 | 1.6 | 1.2 | 11.61 | 159 | 72 | 1.32 | 1.3 | 50 |
| MWC 758 | -7.35 | 1.8 | 0.9 | 8.2 | 160 | 62 | 1.77 | 14 | 10 |
| PDS 70 | -10.22 | 1.3 | 0.2 | 11.71 | 113 | 76 | 0.8 | 0.3 | 49 |
| UX Tau A | -7.95 | 1.7 | 1.6 | 9.83 | 140 | 33 | 1.4 | 2.5 | 40 |
| V1247 Ori | -8.0 | 1.4 | 0.1 | 9.8 | 400 | 64 | 1.82 | 15 | 50 |
| AA Tau | -8.44 | 1.3 | 2.4 | 11.0 | 137 | 44 | 0.68 | 1.1 | 60 |
| AB Aur | -6.8 | 1.7 | 0.7 | 6.9 | 163 | 156 | 2.56 | 65.1 | 40 |
| CQ Tau | ~ -10.5 | 1.6 | 1.3 | 9.0 | 163 | 50 | 1.63 | 10 | 30 |
| CS Cha | -8.3 | 1.6 | 1.0 | 11.1 | 176 | 37 | 1.4 | 1.9 | 20 |
| DM Tau | -8.3 | 1.7 | 0.55 | 13.1 | 145 | 25 | 0.39 | 0.2 | 40 |
| DoAr 44 | -8.2 | 1.4 | 2.9 | 12.1 | 146 | 40 | 1.4 | 1.9 | 20 |
| GM Aur | -8.3 | 1.6 | 0.3 | 11.7 | 160 | 40 | 1.01 | 1.0 | 60 |
| HD 34282 | ~ -10.5 | 1.4 | 0.2 | 9.77 | 312 | 87 | 2.11 | 10.8 | 70 |
| HD 97048 | ~ -10.5 | 1.5 | 0.9 | 8.3 | 185 | 63 | 2.17 | 30.0 | 60 |
| HP Cha | -8.97 | 1.5 | 1.5 | 12.11 | 160 | 50 | 0.95 | 2.4 | 40 |
| IP Tau | -8.14 | 1.7 | 1.45 | 12.46 | 131 | 25 | 0.54 | 0.6 | 60 |
| RY Lup | -8.2 | 1.7 | 0.7 | 11 | 159 | 69 | 1.4 | 1.9 | ~80 |
| RY Tau | -7.1 | 1.7 | 2.2 | 9.67 | 175 | 27 | 2.25 | 15 | ~80 |
| T Cha | -8.4 | 1.4 | 2.0 | 12.75 | 107 | 34 | 1.12 | 1.3 | ~80 |

Table data from FVDM unless otherwise noted in the text, except for $R_A$ which is extrapolated from Simbad, the planet radius $R_{p3}$ (from the COND models), and the disk inclination is estimated from the images of FVDM.

**Table 3:** The Predicted Planetary Parameters from the MAG Model of Gap Planets

| Name | Orbital semi-major axis (au) | | | Average Projected* separation on-sky (") | | | Planet/Star contrast at $\Delta$magH$\alpha$ (mag)** | | | Predicted Mass of planet ($M_{jup}$)** | | |
|---|---|---|---|---|---|---|---|---|---|---|---|---|
| | $a_1$ | $a_2$ | $a_3$ | $Sep_1$ | $Sep_2$ | $Sep_3$ | $\Delta H\alpha_1$ | $\Delta H\alpha_2$ | $\Delta H\alpha_3$ | $Mp_1$ | $Mp_2$ | $Mp_3$ |
| HD 100453 | 8.97 | 14.23 | 22.59 | 0.08 | 0.13 | 0.21 | 11.11 | **11.61** | **12.12** | 5.88 | 2.94 | 1.47 |
| HD 100546 | 8.67 | 13.76 | 21.84 | 0.07 | 0.11 | 0.17 | 7.42 | 7.92 | 8.43 | 8.52 | 4.26 | 2.13 |
| HD135344B | 15.54 | 24.67 | 39.16 | 0.11 | 0.18 | 0.29 | 9.03 | 9.54 | 10.04 | 6.04 | 3.02 | 1.51 |
| HD 169142 | 7.77 | 12.33 | 19.58 | 0.07 | 0.11 | 0.17 | 7.86 | 8.37 | 8.87 | 6.6 | 3.3 | 1.65 |
| LkCa 15 | 21.52 | 34.16 | 54.22 | 0.12 | 0.19 | 0.30 | 5.50 | 6.00 | 6.51 | 5.28 | 2.64 | 1.32 |
| MWC 758 | 18.53 | 29.41 | 46.69 | 0.11 | 0.18 | 0.29 | 8.87 | 9.37 | 9.88 | 7.08 | 3.54 | 1.77 |
| PDS 70 | 22.71 | 36.05 | 57.23 | 0.18 | 0.28 | 0.45 | 7.25 | **8.44** | **8.94** | 3.20 | 1.6 | 0.80 |
| UX Tau A | 9.86 | 15.65 | 24.85 | 0.06 | 0.10 | 0.16 | 6.98 | 7.49 | 7.99 | 5.60 | 2.8 | 1.40 |
| V1247 Ori | 19.13 | 30.36 | 48.19 | 0.04 | 0.07 | 0.10 | 7.54 | 8.05 | 8.55 | 7.28 | 3.64 | 1.82 |
| AA Tau | 13.15 | 20.87 | 33.13 | 0.08 | 0.12 | 0.2 | 7.72 | 8.22 | 8.73 | 2.72 | 1.36 | 0.68 |
| AB Aur | 46.62 | 74.00 | 117.47 | 0.26 | 0.41 | 0.66 | 8.68 | 9.18 | 9.69 | 10.2 | 5.12 | 2.56 |
| CQ Tau | 14.94 | 23.72 | 37.65 | 0.09 | 0.14 | 0.22 | 11.96 | **13.48** | **13.98** | 6.52 | 3.26 | 1.63 |
| CS Cha | 11.06 | 17.55 | 27.86 | 0.06 | 0.1 | 0.15 | 6.15 | 6.66 | 7.16 | 5.60 | 2.8 | 1.40 |
| DM Tau[+] | 7.47 | 11.86 | 18.82 | 0.05 | 0.07 | 0.12 | 7.36[+] | 7.86[+] | 8.36[+] | 1.56 | 0.78 | 0.39 |
| DoAr 44[+] | 11.95 | 18.98 | 30.12 | 0.08 | 0.13 | 0.20 | 8.58[+] | 9.09[+] | 9.59[+] | 5.60 | 2.80 | 1.40 |
| GM Aur | 11.95 | 18.98 | 30.12 | 0.06 | 0.10 | 0.15 | 4.88 | 5.39 | 5.89 | 4.04 | 2.02 | 1.01 |
| HD 34282 | 26.00 | 41.27 | 65.51 | 0.06 | 0.10 | 0.16 | 11.21 | 11.72 | **12.09** | 8.44 | 4.22 | 2.11 |
| HD 97048 | 18.83 | 29.89 | 47.44 | 0.08 | 0.13 | 0.21 | 12.28 | **12.68** | **13.04** | 8.68 | 4.34 | 2.17 |
| HP Cha[+] | 14.94 | 23.72 | 37.65 | 0.09 | 0.14 | 0.21 | 8.79[+] | 9.30[+] | 9.80[+] | 3.8 | 1.9 | 0.95 |
| IP Tau[+] | 7.47 | 11.86 | 18.82 | 0.05 | 0.07 | 0.12 | 7.07[+] | 7.57[+] | 8.08[+] | 2.16 | 1.08 | 0.54 |
| RY Lup | 20.62 | 32.73 | 51.96 | 0.08 | 0.13 | 0.21 | 5.61 | 6.11 | 6.62 | 5.6 | 2.8 | 1.40 |
| RY Tau | 8.07 | 12.81 | 20.33 | 0.03 | 0.05 | 0.07 | 6.46 | 6.96 | 7.47 | 9.00 | 4.50 | 2.25 |
| T Cha[+] | 10.16 | 16.13 | 25.6 | 0.06 | 0.1 | 0.15 | 6.65[+] | 7.16[+] | 7.66[+] | 4.48 | 2.24 | 1.12 |

*We note that this is simply an average position, the true position on the sky depends on the unknown orbital phase and so these *sep* values can underestimate the true *sep* by $(a/\pi D)(\pi-2)(1-\cos(inclination))$ and overestimate by $(a/D)[\cos(inclination)-(1+((2-\pi)/\pi)(1-\cos(inclination)))]$ arcsec.

** Assuming $Mp_1=2Mp_2$ and $Mp_2=2Mp_3$. The $\Delta H\alpha_1$ contrasts could have errors of up to 1.0 mag to -0.6 mag and $\Delta H\alpha_2$ contrasts could have errors of +0.5 mag to -0.3 mag if the mass ratios vary from 1.4x to 3x instead of 2x. Values in **Bold** text are weak accretors and have $\Delta$magH$\alpha$ calculated by equation 9 (all others use equation 8).

[+] faint $R_A$ >12 mag AO targets have had their contrasts increased by +2mag so they can be compared to the AO sensitivity limits in Fig 8. If they were observed from space -2 mag should be applied to get correct contrast.